\documentclass[aps,pre,twocolumn,showpacs]{revtex4-1}
\usepackage{amsmath,graphicx,amsbsy,amssymb,amsmath,epsfig,latexsym}
\usepackage{wasysym,pifont,mathrsfs,float,color,comment,verbatim,multirow,array}

\begin{document}
\title{Sandpile model on Scale Free Networks with preferential sand
  distribution: a new universality class}

 \author{Himangsu Bhaumik}
 \author{S. B. Santra} \email{santra@iitg.ac.in}
 \affiliation{Department of Physics, Indian Institute of Technology
   Guwahati, Guwahati-781039, Assam, India.}  \date{\today}

\begin{abstract}
 A two state sandpile model with preferential sand distribution is
 developed and studied numerically on scale free networks with
 power-law degree ($k$) distribution, {\em i.e.}: $P_k\sim
 k^{-\alpha}$. In this model, upon toppling of a critical node sand
 grains are given one to each of the neighbouring nodes with highest
 and lowest degrees instead of two randomly selected neighbouring
 nodes as in a stochastic sandpile model. The critical behaviour of
 the model is determined by characterizing various avalanche
 properties at the steady state varying the network structure from
 scale free to random, tuning $\alpha$ from $2$ to $5$. The model
 exhibits mean field scaling on the random networks,
 $\alpha>4$. However, in the scale free regime, $2<\alpha<4$, the
 scaling behaviour of the model not only deviates from the mean-field
 scaling but also the exponents describing the scaling behaviour are
 found to decrease continuously as $\alpha$ decreases. In this regime,
 the critical exponents of the present model are found to be different
 from those of the two state stochastic sandpile model on similar
 networks. The preferential sand distribution thus has non-trivial
 effects on the sandpile dynamics which leads the model to a new
 universality class.
\end{abstract}
%\pacs{complex-system,networks,soc,crack-sandpile-avalanche}
\maketitle

\section{Introduction} 
The degree distribution $P_k$, probability to have a node with degree
$k$, of a class of complex networks \cite{naim,newman} is given by
\begin{equation}
  P_k\sim k^{-\alpha} 
\label{Pk}
\end{equation}
where $\alpha$ is the characteristic degree exponent. Such networks
are known as scale-free network (SFN) \cite{albertREVMP02}. The
behaviour of many systems like epidemic spreading
\cite{satorrasPRL01,mayPRE01}, percolation \cite{cohenPRE02},
etc. occurring on the SFN have strong dependence on $\alpha$ because
of diverging moment of degree distribution; $\langle
k^n\rangle\to\infty$ for $n\ge\alpha-1$ as $k_{\rm max}$ is infinitely
large. Depending upon the value of $\alpha$ the network has two
regimes scale free and random. The scale free regime, $2\le \alpha \le
3$, is characterized by finite average degree $\langle k\rangle$ and
diverging second moment of the degree distribution $\langle
k^2\rangle$ whereas in the random regime both are found to be finite
\cite{barabasi}. Furthermore, due to strong heterogeneity in the
degree distribution, the translational symmetry, nodes with similar
neighbourhood, is absent in the scale free regime whereas it is a
necessary condition in the random regime \cite{wuPRE07}. The
percolation model \cite{cohenPRE02} as well as the epidemic spreading
model \cite{wangRPP17} exhibits mean-field (MF) behaviour on random
network for $\alpha>4$ whereas shows some non-trivial critical
behaviour other than MF on scale free network for $\alpha<4$. On the
other hand, due to the heterogeneity in degrees, SFN with $\alpha\le3$
are highly resilience against random attack \cite{rekaNAT00} whereas
it is vulnerable when the attack is targeted on a few nodes with
larger degree \cite{gallosPRL05}. Under such intentional attack a
devastating cascading failure such as black out of electric power grid
in an entire country could happen through out a network
\cite{motterPRE02,sachtjenPRE00}.

Avalanche dynamics of a sandpile model \cite{bak,jensen,pruessner}
manifests such cascading effects where the system in the critical
state is triggered by a small perturbation and the response spreads
all over the system redistributing sand (or energy) in a cascaded
manner. Various types of sandpile models in Euclidean space have been
studied introducing different kinds of constraints in the sand
distribution dynamics via well defined toppling rules such as
stochastic \cite{mannaPHYA91}, directional \cite{dharPRL89},
rotational \cite{santraPRE07} etc. In all such cases, the models are
found to belong to different universality classes. Both Bak Tang
Wiesenfeld (BTW) sandpile model \cite{btwPRL87,btwPRA88} and
stochastic sandpile model (SSM) \cite{mannaPHYA91} have been
introduced on SFN \cite{gohPRL03,leePHYA04,*gohPHYA05}. Though the
avalanche size distribution is highly affected by $\alpha$ in the BTW
model, it represents MF behaviour for the SSM for all values of
$\alpha\ge2$. However, to our knowledge, sandpile model on SFN with
targeted sand distribution to nodes with specific degrees is still not
reported. Since the targeted sand distribution in certain models leads
to non-trivial scaling behaviour, it is intriguing to develop a
sandpile model on SFN distributing sand to nodes with specific degrees
and study the sandpile dynamics on varied scale free structures.

It this paper, a two state sandpile model with preferential sand
distribution, in short a preferential sandpile model (PSM), is
developed and studied on SFN varying the degree exponent $\alpha$. In
the preferential rule, sands are given to the highest and the lowest
degree neighbouring nodes in the event of toppling of a critical
node. The model represents nontrivial critical behaviour for
$\alpha<4$ and MF behaviour for $\alpha > 4$ as seen in the epidemic
spreading model on SFN \cite{wangRPP17}. Results are compared with
those of the SSM, a two state stochastic sandpile model, on SFN
\cite{gohPRL03}.

\section{The Model}
\begin{figure}[t]
\centerline{\hfill
\psfig{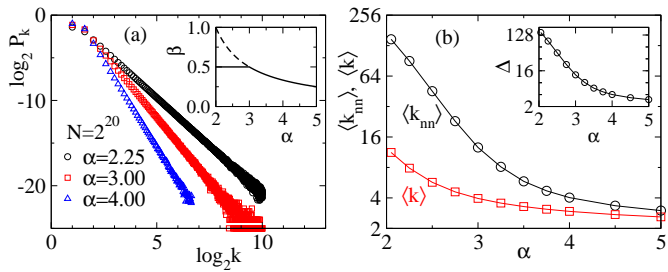}\hfill
  }
\caption{\label{Pk_knn}(Colour online) (a) Plot of $P_k$ against $k$
  for different values of $\alpha$ for a network of size
  $N=2^{20}$. The distribution is sampled over $32$
  configurations. Inset: Plot of $\beta=\ln{k_{\rm max}}/\ln{N}$ $vs$
  $\alpha$ shown in solid line. The dashed line shows the natural
  cutoff $\beta=1/(\alpha-1)$ for $\alpha<3$. (b) Plot of $\langle
  k_{nn} \rangle$ ($\Circle$) and $\langle k \rangle$ ($\Box$) against
  $\alpha$. Inset shows how their difference $\Delta=\langle k_{nn}
  \rangle - \langle k \rangle$ changes with $\alpha$.}
\end{figure}

Scale free networks of $N$ nodes are generated employing uncorrelated
configurational model \cite{catanzaroPRE05}. In order to get a
scale-free degree distribution, a random number $r$, uniformly
distributed between [$0,1$], is drawn for each node and degree $k={\rm
  INT}[k_{\rm min}/r^{1/(\alpha-1)}]$ where $k_{\rm min}=2$, is
assigned. The natural upper cutoff of the degree for a SFN is $k_{\rm
  max}=N^\beta$ where $\beta=1/(\alpha-1)$. However, for $\alpha\le 3$
the upper cutoff is set to $\sqrt{N}$ instead of the natural cutoff
which eventually would satisfy the conditions of no multi-edge or
self-edge of the nodes in a uncorrelated random SFN.  The degree
distributions of some of the networks considered are shown in
Fig. \ref{Pk_knn}(a) for different values of $\alpha$. The values of
$\beta$ used to fix $k_{\rm max}$ for different values of $\alpha$ are
shown in the inset of Fig. \ref{Pk_knn}(a). The likely neighbourhood
of nodes is identified by comparing the average nearest neighbour
degree $\langle k_{nn}\rangle$ with the average degree $\langle
k\rangle$ of a network as shown in Fig. \ref{Pk_knn}(b). The
difference between the two averages $\Delta=\langle k_{nn}\rangle -
\langle k\rangle$ is shown in the inset of Fig. \ref{Pk_knn}(b) and
found to be quite large for small $\alpha$ and decreases to $k_{\rm
  min}$ as $\alpha \rightarrow \infty$.

\begin{figure}[t]
\centerline{\hfill
  \psfig{file=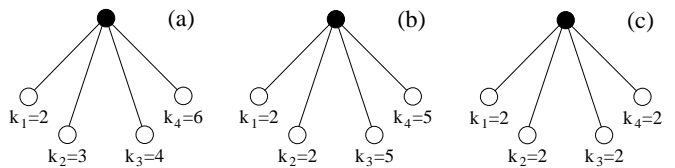,width=0.49\textwidth}\hfill}
\caption{\label{trule} Toppling rules for PSM on a node of degree
  $k=4$ are demonstrated. The black filled circle represents the
  critical node which has $4$ adjacent nodes (open circle) with
  degree, say, $k_1,k_2,k_3,$ and $k_4$. In (a) $k_1<k_2<k_3<k_4$, the
  nodes with degree $k_1$ and $k_4$ receive one sand grain each. In
  (b) $k_1=k_2<k_3=k_4$, two sand grains are distributed among one of
  the randomly chosen node from $\{k_1,k_2\}$ and another from
  $\{k_3,k_4\}$. In (c) $k_1=k_2=k_3=k_4$, two sand grains are
  randomly given to any two distinct nodes.}
\end{figure}

In order to implement sandpile model on a network where there is no
boundary, one needs to have an estimate for the rate of dissipation of
sand grains during its flow over the network. Bulk dissipation is one
of the options in which a sand grain is removed from the system with a
small probability $\epsilon$ during the transfer of a sand grain from
one node to another. For given $\alpha$ and $\epsilon$, the SFN with
$N$ nodes is driven by adding sand grains, one at a time, to a
randomly chosen node $i$. If the height $h$ of the sand column at the
$i$th node is greater than or equal to the threshold value $h_c=2$,
the sand column becomes unstable or critical and collapses by
distributing two sand grains to the neighbouring nodes with the highest
and lowest degree among the $k_i$ adjacent nodes. If more than one
node has the same highest (or lowest) degree, one of them is chosen
randomly to consider as the highest (or the lowest) degree node. Such
special situations are demonstrated in Fig. \ref{trule}. The toppling
rule of the $i$th critical node of an SFN is then given by
\begin{equation}
\label{trule2}
\begin{array}{ll}
& h_i \rightarrow h_i-h_c, \\ {\rm and} & h_j=
  \left\{\begin{array}{ll} h_j  & \mbox{if} \hspace{0.2cm} r \le
  \epsilon, \\ h_j + 1 & \mbox{otherwise} \end{array}\right.
\end{array}
\end{equation}
where $j$ corresponds to the adjacent nodes with highest and lowest
degree, $r$ is a random number uniformly distributed over $[0,1]$. If
the toppling causes any of adjacent nodes critical, subsequent
toppling follow on these nodes in parallel until all the nodes in the
network become under critical. These toppling activities lead to an
avalanche. As an avalanche seized, another sand grain is added to the
system.

The model is studied varying $\alpha$ from $2$ to $5$ for a given
$\epsilon$. Since on the random network, $\epsilon \sim 1/\sqrt{N}$
\cite{bhaumikPRE13} for a network of $N$ nodes, the dissipation factor
$\epsilon=1/\sqrt{N}$ is taken for most of the values of $\alpha$ and
the effect of $\epsilon$ on the sandpile dynamics is verified for a
few specific values of $\alpha$.

\section{Numerical simulation}
An extensive computer simulation is performed varying $\alpha$ from
$2$ to $5$. For a given $\alpha$, the size of the networks $N$ varied
from $N=2^{16}$ to $N=2^{20}$ in multiple of $2$. In order to estimate
the avalanche properties, the following statistical averages are
made. For a given $\alpha$ and $N$, thirty two different SFN
configurations are considered. On each SFN, $10^6$ avalanches are
collected neglecting the first $3\times10^6$ avalanches during which
steady state has been achieved in all networks. Therefore, for a given
$\alpha$ and $N$, a total of $32\times 10^6$ avalanches is taken for
data averaging. The critical behavior of different avalanche
properties like the toppling size $s$ (total number of toppling in an
avalanche), area $a$ (the number of distinct nodes toppled in an
avalanche), and lifetime $t$ (the number of parallel updates to make
all the nodes under critical) of an avalanche are measured in the
steady state to characterize the PSM on SFN. 

\begin{figure}[t]
\centerline{\hfill
  \psfig{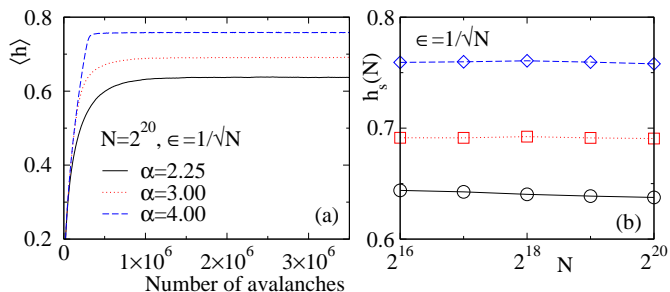}
  \hfill }
\caption{\label{avht}(Colour online) (a) Plot of $\langle h\rangle$
  against number of avalanches on SFN of size $N=2^{20}$ for
  $\alpha=2.25$ (black solid line), $3$ (red dotted line), and $4$
  (blue dashed line) taking $\epsilon=1/\sqrt{N}$. (b) The variation
  of $h_s$ with network size $N$ for $\alpha=2.25$ ($\Circle$), $3$
  ($\Square $), and $4$ ($\Diamond$) for the same value of
  $\epsilon$. }
\end{figure} 

The steady state corresponds to balance of incoming and outgoing
fluxes of sand grains which leads to time independent average height
of the sand columns in the network. For a given $\alpha$, the average
height $\langle h\rangle$ is calculated as
\begin{equation}
\label{st} 
\langle h \rangle = \frac{1}{N}\sum_{i=1}^{N}h_i
\end{equation}  
where $h_i$ is the height of the sand column at the $i$th node and $N$
is the total number of nodes in the network. In Fig. \ref{avht}(a),
$\langle h\rangle$ is plotted against the number of avalanches for
different values of $\alpha$ on a network of size $N=2^{20}$. Starting
from an empty configuration, the steady state of PSM is achieved after
more than $10^6$ avalanches for all values of $\alpha$. It can be seen
that the values of $\langle h\rangle$ at the steady state increases as
$\alpha$ increases. Since the critical height $h_c=2$ in this model,
at the end of an avalanche, the nodes either will have a sand or they
will remain empty. Thus for $\alpha=4$, nearly $75\%$ of the nodes are
having sands whereas for $\alpha=2.25$, nearly $60\%$ of the nodes are
having sands. For a given value of $\alpha$, the saturated average
height $h_s$ of the sand columns in the steady state is estimated
taking average over last $10^5$ avalanches of every $32$ different
configurations. For a given value of $\alpha$, $h_s$ is found to be
independent of network size $N$ as shown in the Fig. \ref{avht}(b) for
three different values of $\alpha$. It has also been verified that
$h_s$ increases if the dissipation factor $\epsilon$ decreases and
vice verse for a given $\alpha$ as expected.

In order to compare the results of PSM with that of the SSM, the above
numerical computation has also been repeated for SSM on the similar
networks. In case of the SSM, similar variations $h_s$ with $\alpha$,
$N$ and $\epsilon$ are observed.

\section{Avalanche evolution}
\begin{figure}[t]
\centerline{\hfill
  \psfig{file=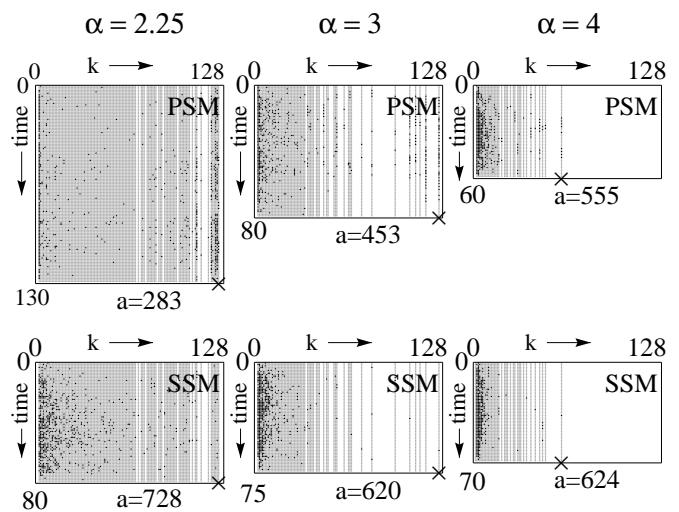,width=0.49\textwidth}
  \hfill}
\caption{\label{morpho} Time evolution of typical avalanches of PSM
  (upper row) and those of SSM (lower row) in degree space are shown
  for $\alpha=2.25$ (left column), $\alpha=3$ (middle column), and,
  $\alpha=4$ (right column) on a network of size $N=2^{14}$ taking
  $\epsilon=1/\sqrt{N}$. The black dots represent the toppled nodes
  and the gray colour represents the nodes with no toppling. The white
  space corresponds to no nodes of such degree. The crosses represent
  the maximum degree present in the network.}
\end{figure} 
Time evolution of a few typical avalanches of PSM generated on a
network of size $N=2^{14}$ are shown in the upper row of
Fig. \ref{morpho} for different values of $\alpha$. For comparison,
time evolved morphology of avalanches in the SSM on the same networks
are given in the lower row of Fig. \ref{morpho}. The degree $k$ of the
nodes is presented along the horizontal axis and time (the parallel
updates) is presented along the downward vertical axis. The black dots
represent the toppled nodes, the gray color corresponds to the nodes
of certain degree with no toppling and white space corresponds to no
node of that degree. The avalanches presented here have a common size
$s=800$ for both the models, the area $a$ are mentioned at the bottom
of each configuration and their lifetime $t$ (maximum number of
parallel updates) can be seen from the vertical axis. The time
evolution of an avalanche of PSM differs considerably than that of SSM
on a given network. For PSM, $a$ increases and $t$ decreases as
$\alpha$ increases whereas for SSM both $t$ and $a$ remains almost
same for all values of $\alpha$. In PSM, small $a$ and large $t$ for
smaller $\alpha$, indicates multiple toppling of the nodes in the
scale free regime ($2\le\alpha\le3$). On the other hand, in the random
regime with $\alpha\ge 4$, large $a$ and small $t$ indicate single
toppling of different nodes. It can also be noted that in PSM the
density of toppled nodes is high at the lower and higher degree nodes
of the network in scale-free regime whereas most of the lower degree
nodes are involved in an avalanche in the random regime. However, in
SSM the density of toppled nodes always decreases with $k$. The
characteristic features of the two models are quite different and
hence it is important to characterize the critical avalanche
properties of PSM quantitatively and compare the results with those of
the SSM on SFN.

\begin{figure}[t]
\centerline{\hfill
  \psfig{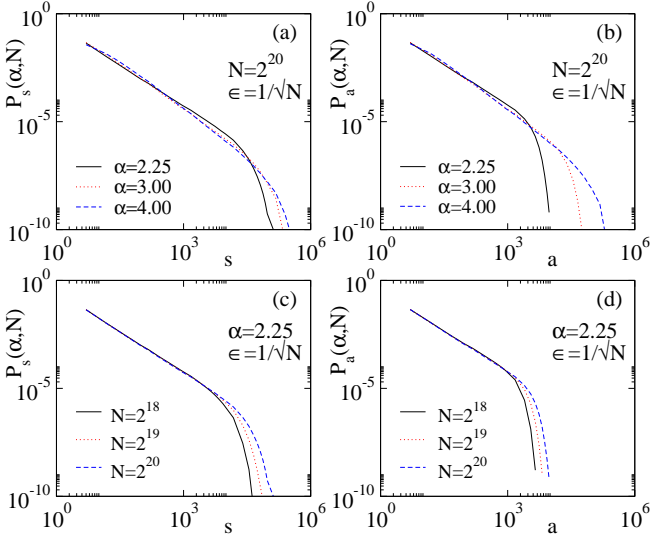}
  \hfill} 
\caption{\label{Ps}(Colour online) Plot of $P_{s}(\alpha,N)$ against
  $s$ and $P_{a}(\alpha,N)$ against $a$ in (a) and (b) respectively
  for $\alpha=2.25$ (solid line), $\alpha=3$ (dotted line), $\alpha=4$
  (dashed line) for a network of size $N=2^{20}$. For a given
  $\alpha$, the distributions for different network sizes $N=2^{18}$
  (solid line), $N=2^{19}$ (dotted line), and $N=2^{20}$ (dashed line)
  are shown in (c) for $P_{s}(\alpha,N)$ and in (d) for
  $P_{a}(\alpha,N)$. All distributions are estimated taking
  $\epsilon=1/\sqrt{N}$. }
\end{figure}

\section{Characterization of PSM on SFN}
 To characterize the properties of PSM, the probability distributions
 $P_{x}(\alpha,N)$ of avalanche properties $x\in\{s,a\}$ at the
 critical steady state are determined for various values of degree
 exponent $\alpha$, dissipation factor $\epsilon$ and network size
 $N$. Distributions will be studied taking $\epsilon=1/\sqrt{N}$ and
 the effect of $\epsilon$ on the distributions will be analyzed later
 for a specific value of $\alpha$.  For a fixed $N=2^{20}$, the
 distributions $P_{s}(\alpha,N)$ and $P_{a}(\alpha,N)$ are plotted in
 Figs. \ref{Ps}(a) and \ref{Ps}(b) respectively for several values of
 $\alpha$. Keeping $\alpha$ fixed at $2.25$, the same distributions
 $P_{s}(\alpha,N)$ and $P_{a}(\alpha,N)$ for different values of $N$
 are plotted in Figs. \ref{Ps}(c) and \ref{Ps}(d) respectively. Though
 the cutoffs depend on both $N$ and $\alpha$ for a given $\epsilon$,
 the scaling exponents seem to be independent of the network size $N$
 for a given $\alpha$ but it depends on $\alpha$ for a given
 $N$. Hence, for given $\alpha$ and $N$, a finite size scaling (FSS)
 form of $P_{x}(\alpha,N)$ is assumed as
\begin{equation}
\label{psN}
P_{x}(\alpha,N) =
x^{-\tau_x(\alpha)}f_{x,\alpha}\left[\frac{x}{N^{D_x(\alpha)}}\right],
\end{equation}
where $\tau_x(\alpha)$ is the scaling exponent, $D_x(\alpha)$ is the
capacity dimension, and $f_{x,\alpha}$ is a $\alpha$ dependent scaling
function of an avalanche property $x$ for a given $\epsilon$. Two
important aspects of the distributions need to be verified. First is
the universality, {\em i.e.} determination of the values of the
exponents and second is the verification of the FSS form
assumed. Though the two state models like SSM
\cite{mannaJPA91,dharPHYA99a} and rotational sandpile model
\cite{santraPRE07} follow FSS on regular lattice, the BTW model does
not. It is then intriguing to verify whether FSS of PSM on SFN is
valid or not.

\begin{figure}[t]
 \centerline{ \hfill 
   \psfig{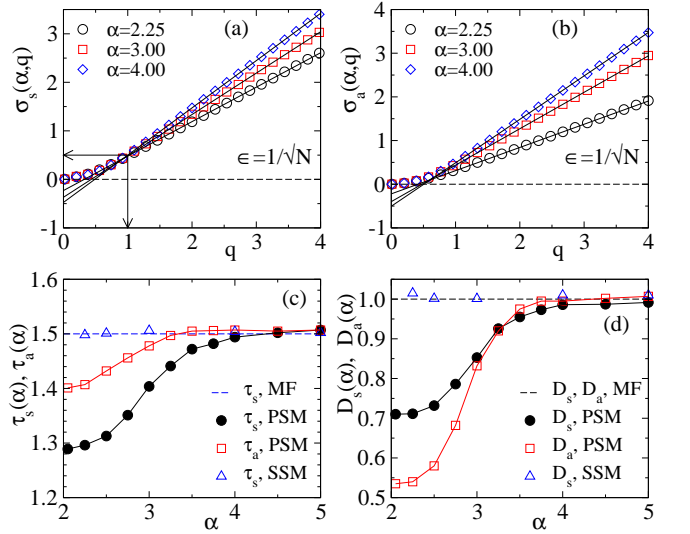}\hfill}
  \caption{\label{sig-tau}(Colour online) Plot of (a)
    $\sigma_s(\alpha,q)$ and (b) $\sigma_a(\alpha,q)$ against $q$ for
    $\alpha=2.25$ ($\Circle$), $\alpha=3$ ($\Box$), and $\alpha=4$
    ($\Diamond$). For clarity only $25$ points out of $400$ points are
    shown. The solid lines represent the linear least square fit
    through the data points. (c) Plot of $\tau_s(\alpha)$ (black
    filled circle) and $\tau_a(\alpha)$ (red open square) against
    $\alpha$. For comparison, $\tau_s(\alpha)=\tau_a(\alpha)$ for SSM
    are given in blue triangles. (d) Plot of $D_s(\alpha)$ (black
    filled circle) and $D_a(\alpha)$ (red open square) against
    $\alpha$. The values of $D_s(\alpha)=D_a(\alpha)$ for SSM are
    shown in blue triangles. The dashed lines in (c) and (d)
    represents the MF value. The error in the values of the exponents
    are of the order of symbol size.}
\end{figure}
In order to estimate the values of the exponents $\tau_x(\alpha)$ and
$D_x(\alpha)$, [Eq. (\ref{psN})], the concept of moment analysis
\cite{lubeckPRE00a} has been used. For a given $\alpha$, the $q$th
moment of $x$ as function of $N$ can be obtained as
\begin{eqnarray}
\label{sq}
\langle x^q(\alpha,N) \rangle &=& \int_0^{\infty}
x^{q}P_{x}(\alpha,N)dx \sim N^{\sigma_{x}(\alpha,q)},
\end{eqnarray}
where the $q$th moment scaling exponent
\begin{equation}
\label{sig}
\sigma_{x}(\alpha,q)=D_x(\alpha)q+D_x(\alpha)[1-\tau_x(\alpha)]
\end{equation}
for $q>\tau_x(\alpha)-1$ and it is zero for $q<\tau_x(\alpha)-1$. For
each value of $\alpha$, a sequence of values of $\sigma_{x}(\alpha,q)$
as a function of $q$ is determined by estimating the slope of the
plots of $\log\langle x^q(\alpha,N)\rangle$ versus $\log(N)$ for $400$
equidistant values of $q$ between $0$ and $4$. $\sigma_{s}(\alpha,q)$
and $\sigma_{a}(\alpha,q)$ are plotted against $q$ for $\alpha=2.25$,
$3$, and $4$ in Figs. \ref{sig-tau}(a) and \ref{sig-tau}(b)
respectively. First, it can be seen that for $q=1$, the value of
$\sigma_{s}(\alpha,1)$ is found to be $\approx 1/2$ (indicated by
arrows in Fig. \ref{sig-tau}(a)) irrespective of the values of
$\alpha$. Since the dissipation factor is taken as
$\epsilon=1/\sqrt{N}$ for all values of $\alpha$, the average number
of toppling required for an avalanche to dissipate one sand grain is
$\sqrt{N}$. Hence, $\langle s \rangle \sim N^{1/2}$,
i.e. $\sigma_{s}(\alpha,1)=1/2$ for all $\alpha$. In order to estimate
the values of the exponents, the direct method developed by L\"ubeck
\cite{lubeckPRE00a} is employed. Following such method, straight lines
are fitted through the data points $\{\sigma_{x}(\alpha,q),q\}$ in the
range of $2\le q\le 4$ and the exponents $\tau_x(\alpha)$ and
$D_x(\alpha)$ are obtained from the intercepts on the $q$-axis and
$\sigma_x(\alpha,q)$ axis respectively for a given $\alpha$. Following
Eq. (\ref{sig}), the $q$ intercept provides $\tau_x(\alpha)-1$, the
$\sigma_x(\alpha,q)$ intercept provides
$D_x(\alpha)[1-\tau_x(\alpha)]$. The estimated values of the exponents
are: $\tau_s=1.296(7)$, $\tau_a=1.407(6)$, $D_s=0.720(6)$,
$D_a=0.534(5)$ for $\alpha=2.25$; $\tau_s=1.403(6)$,
$\tau_a=1.478(6)$, $D_s=0.852(5)$, $D_a=0.837(6)$ for $\alpha=3$;
$\tau_s=1.491(5)$, $\tau_a=1.503(4)$,$D_s=0.975(6)$,$D_a=0.994(5)$ for
$\alpha=4$. The number in the parentheses is the uncertainty of last
digit in the numerical value of the respective exponents. The values
of $\tau_x(\alpha)$ and $D_x(\alpha)$ are estimated at various
different values of $\alpha$ and presented as a function of $\alpha$
in Figs. \ref{sig-tau}(c) and \ref{sig-tau}(d) respectively.

In order to compare the values of the exponents of PSM with those of
the SSM, the estimates of the values of these exponents for the SSM
are also presented in the same figures. The estimated exponents are
not only found to be the same as reported in \cite{gohPRL03} but also
same as that of MF \cite{bonabeau95,christensenPRE93}. It is already
reported that the SSM exhibits MF behaviour \cite{gohPRL03} throughout
the range of $\alpha$, scale free as well as random. It is also known
that the SSM has a different scaling behaviour than MF on SWN
\cite{bhaumikPRE16} in contrary to the present observation. It is
worth mentioning here that for BTW type deterministic sandpile model
on SFN, the exponent $\tau_s(\alpha)$ also has a continuous dependence
on $\alpha$ as $\tau_s(\alpha)=\alpha/(\alpha-1)>3/2$ in the range
$2<\alpha<3$ and remains $\tau_s(\alpha)=3/2$ for $\alpha>3$
\cite{gohPRL03}. However, in PSM, all four exponents, $\tau_s$,
$\tau_a$, $D_s$, $D_a$, have MF values in the random regime
($\alpha\ge 4$) but they vary continuously with $\alpha$ but remain
lower than the MF values in the scale free regime ($\alpha\le3$). In
BTW, the nodes with higher degree sustain large number of sand grains
and play a role of reservoirs whereas in PSM no nodes of any degree
sustain large number of sand grains, hence, such reservoirs do not
exist. On the other hand, in MF analysis loop less structures in the
branching process, nodes without multiple toppling, are
assumed. Hence, the distributions $P_s$ and $P_a$ are characterized by
the same exponents $\tau_s=\tau_a$ and $D_s=D_a$. Thus the network
structure (scale free or random) plays a crucial role in determining
the critical behaviour of PSM in contrary to SSM.

\begin{comment}
From the time evolution morphology of an avalanche (see
Fig. \ref{morpho}), for SSM, the avalanche area ($a$) is found to be
almost same as that of the avalanche size ($s$) at all values of
$\alpha$; whereas for PSM, $a$ is much less than $s$ at lower $\alpha$
and it becomes comparable with $s$ for $\alpha\ge4$. This explains
qualitatively the existence of uniform scaling behaviour of SSM
throughout the range of $\alpha$ whereas an $\alpha$ dependent scaling
behaviour in PSM.
\end{comment}

\begin{figure}[t]
  \centerline{\hfill
  \psfig{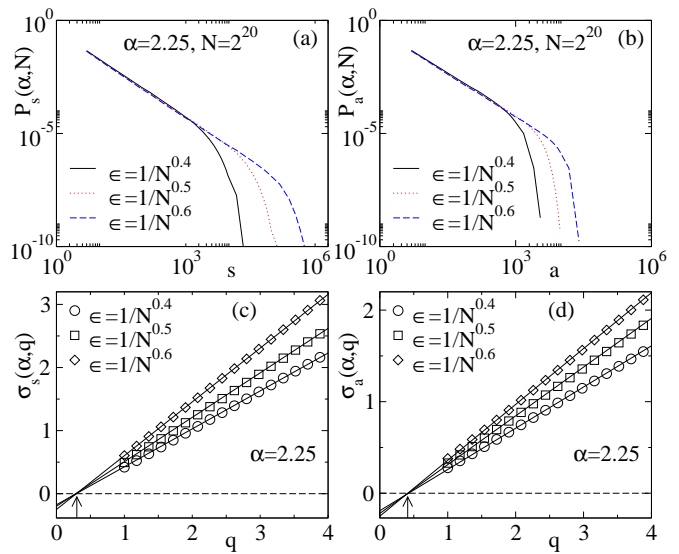}
  \hfill}
\caption{\label{sig-eps}(Colour online) Plot of (a) $P_s(\alpha,N)$
  and (b) $P_a(\alpha,N)$ for $\epsilon=1/N^{0.4}$ (black solid line),
  $\epsilon=1/N^{0.5}$ (red dotted line), and $\epsilon=1/N^{0.6}$
  (blue dashed line) for a fixed value of $\alpha=2.25$ and
  $N=2^{20}$. Plot of (c) $\sigma_s(\alpha,q)$ and (d)
  $\sigma_a(\alpha,q)$ against $q$ for $\epsilon=1/N^{0.4}$
  ($\Circle$), $\epsilon=1/N^{0.5}$ ($\Box$), and $\epsilon=1/N^{0.6}$
  ($\Diamond$). The solid lines represent the linear least square fit
  through the data points. The intersection points of the fitted lines
  are marked by arrow heads on the $q$-axis. The dashed lines
  correspond to $\sigma_x=0$. For clarity points for $q<1$ are also
  dropped.}
\end{figure}

It is now important to verify the effect of the choice of $\epsilon$
on the scaling behaviour of PSM for a given $\alpha$. The
distributions $P_s(\alpha,q)$ and $P_a(\alpha,q)$ for three different
values of $\epsilon$ are obtained and shown in Figs. \ref{sig-eps}(a)
and \ref{sig-eps}(b) respectively for $\alpha=2.25$. The cutoff of the
distributions are found to depend on $\epsilon$ as expected. Estimates
of $\sigma_s(\alpha,q)$ and $\sigma_a(\alpha,q)$ are made for all
three values of $\epsilon$ at $\alpha=2.25$. Variation of
$\sigma_s(\alpha,q)$ and $\sigma_a(\alpha,q)$ against $q$ are shown in
Figs. \ref{sig-eps}(c) and \ref{sig-eps}(d) respectively. First of all
the value of $\sigma_s(1)$ for $q=1$ are found to increases as the
value of $\epsilon$ decreases as expected. Secondly, the plots
intersect the $q$-axis at a single point. Since the $q$ intercepts are
$\tau_s-1$ or $\tau_a-1$, the scaling exponents $\tau_s$ or $\tau_a$
remain independent of the choice of $\epsilon$. For $\alpha=2.25$, the
estimated values of $\tau_s$ and $\tau_a$ for $\epsilon=1/N^{0.4}$ and
$1/N^{0.6}$ are found to be within the error bar of the corresponding
value of $\tau_s$ and $\tau_a$ for $\epsilon=1/N^{0.5}$. However, the
slope of the plots are found strongly dependent on $\epsilon$. Since
the slope determines the capacity dimension $D_x$, it should depend on
$\epsilon$ for a given $\alpha$. For $\alpha=2.25$, the values of
$D_s$ are found to be $0.610(5)$ and $0.857(4)$ for
$\epsilon=1/N^{0.4}$ and $1/N^{0.6}$ respectively and are out of the
error bars of the corresponding value of $D_s=0.720(6)$ for
$\epsilon=1/N^{0.5}$. A similar result is also observed for
$D_a$. Such dependence of the critical exponents on the choice of the
dissipation factor is also reported in few other studies
\cite{gohPRL03,karmakarJPA05}.

It can be noted here that the results of PSM obtained here on the
uncorrelated SFNs. The results have also been verified for correlated
Barabasi-Albert SFN with $\alpha=3$, generated by preferential
attachment method \cite{barabasiPHYA99,*barabasiSC99} and the
distributions $P_s$ and $P_a$ are found similar to those of PSM on the
corresponding uncorrelated SFN with $\alpha=3$. However, the SSM on
optimized Barabasi-Albert SFN imposed on two dimensional square
lattice with degree exponent $\alpha=3$ exhibits scaling behaviour
with exponent $\tau_s=1.30$ \cite{karmakarJPA05}. A similar result is
also obtained in the study of BTW type sandpile model on geographically
embedded SFNs \cite{huangPRE06}. This is because of the fact that the
optimization process destroys the small-world behavior though the
degree distribution remains scale-free.

The scaling behaviour of $P_x(\alpha,N)$ are found to be independent
on the choice of $k_{\rm max}$, the cutoff degree of the network. It
is also reported in the recent study of explosive percolation on SFN
that the choice of cutoff degree of a network has little influence on
the scaling behaviour of geometrical quantities
\cite{radicchiPRL09}.

\begin{figure}[t]
\centerline{\hfill
\psfig{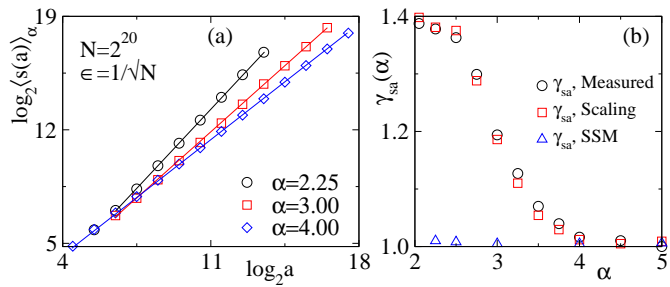}
 \hfill}
\caption{\label{CEsa}(Colour online) (a) Plot of $\langle
  s(a)\rangle_\alpha$ against $a$ on a network of size $N=2^{20}$ with
  $\epsilon=1/\sqrt{N}$ for $\alpha=2.25$ ($\Circle$), $\alpha=3$
  ($\Box$), and $\alpha=4$ ($\Diamond$). (b) Plot of
  $\gamma_{sa}(\alpha)$ versus $\alpha$; Circles represent the
  measured values of $\gamma_{sa}$ and squares represent the same
  obtained from Eq. (\ref{gama-tau}). Blue triangles in (b) represent
  $\gamma_{sa}$ for the SSM. }
\end{figure}
Further insight can be obtained by studying the conditional
expectation $\langle s(a)\rangle_\alpha$ of the avalanche size $s$ for
a fixed area $a$. For a given $\alpha$, $\langle s(a)\rangle_\alpha$
expected to scale as 
\begin{equation}
\label{gamasa}
\langle s(a)\rangle_\alpha = \int sP(s|a) ds \sim a^{\gamma_{sa}(\alpha)},
\end{equation}
where $P(s|a)$ is the conditional probability distribution and
$\gamma_{sa}(\alpha)$ is an exponent. The exponent
$\gamma_{sa}(\alpha)$ is expected to satisfy a scaling relation
\begin{equation}
\label{gama-tau}
\gamma_{sa}(\alpha)=\frac{\tau_a(\alpha)-1}{\tau_s(\alpha)-1}
\end{equation}
with the exponents $\tau_s(\alpha)$ and $\tau_a(\alpha)$ as in usual
sandpile models \cite{jensen}. The exponent $\gamma_{sa}(\alpha)$ is
now measured and the scaling relation is verified. In
Fig. \ref{CEsa}(a), $\langle s(a)\rangle_\alpha$ is plotted against
$a$ in double logarithmic scale for three different values of $\alpha$
taking $\epsilon=1/\sqrt{N}$. It can be seen that for a given
$\alpha$, $\langle s(a)\rangle_\alpha$ scales with $a$ as given in
Eq. (\ref{gamasa}). Obtaining the slope by linear least square method
through the data points, the values of $\gamma_{sa}$ are measured and
they are found as $1.39\pm 0.01$, $1.19\pm 0.01$, and $1.01\pm 0.01$
for $\alpha=2.25$, $3$, and $4$ respectively. The values of the
exponent $\gamma_{sa}(\alpha)$ are also measured for other values of
$\alpha$ and its variation with $\alpha$ is shown in the
Fig. \ref{CEsa}(b). For $\alpha\ge 4$, not only the exponent
$\gamma_{sa}\approx 1$ but also the avalanche size $s$ is equal to
area $a$. This indicates that the nodes toppled only once during the
avalanche and that is why the value of $\tau_s$ is that of
MF. Whereas, for $\alpha<4$, the value of $\gamma_{sa}$ is more than
one and found to be $\approx 1.4$ as $\alpha\to 2$. Not only the value
of $\gamma_{sa}$ is higher at a smaller $\alpha$ than at a higher
$\alpha$, the absolute value of avalanche size $s$ is found to be
higher for smaller values of $\alpha$ at given avalanche area
$a$. Thus, in the scale free regime ($\alpha<4$), a single node must
have toppled multiple times to have higher avalanche size keeping
avalanche area fixed. In order to verify the scaling relation given in
Eq. (\ref{gama-tau}), the estimates of
$(\tau_a(\alpha)-1)/(\tau_s(\alpha)-1)$ are also plotted in
Fig. \ref{CEsa}(b) and compared with the directly measured value of
$\gamma_{sa}$. It can be seen that the scaling holds within the error
bars. As the exponents $\tau_s$ and $\tau_a$ are independent of
dissipation factor $\epsilon$, the exponent $\gamma_{sa}$ is also
independent of $\epsilon$.

As the values of the critical exponents are found very different from
those of the SSM in the scale free regime ($\alpha<4$), PSM then
belongs to a new universality class than SSM in this regime of
SFN. The results of the above model remain unchanged even if the two
sand grains of a critical node goes randomly and independently to any
of the highest and the lowest degree neighbouring nodes instead of
giving one sand grain each to the highest and the lowest degree
neighbouring nodes.

\begin{figure}[t]
\centerline{\hfill
  \psfig{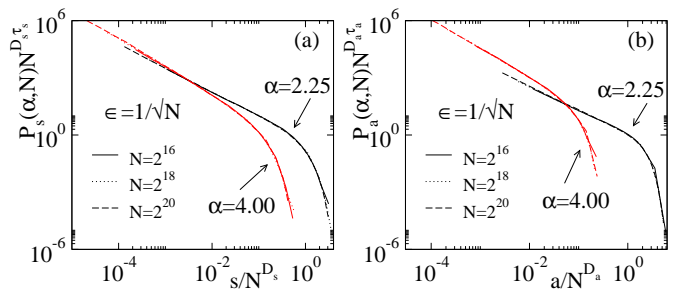}
  \hfill}
\caption{\label{DC}(Colour online) Plot of
  $P_x(\alpha,N)N^{D_x(\alpha)\tau_x(\alpha)}$ $vs$
  $x/N^{D_x(\alpha)}$ for $N=2^{16}$ (solid line), $N=2^{18}$ (dotted
  line), $N=2^{20}$ (dashed line) for $x=s$ in (a) and for $x=a$ in
  (b) taking corresponding values of exponents. Distributions for
  $\alpha=2.25$ and for $\alpha=4$ are marked by arrows.}
\end{figure}
Knowing the values of the exponents $\tau_x$ and $D_x$ for a given
$\alpha$ and $\epsilon$, the scaling function form of $P_x(\alpha,N)$
given in Eq. (\ref{psN}) is verified for both $s$ and $a$. The scaled
avalanche size distribution
$P_s(\alpha,N)N^{D_s(\alpha)\tau_s(\alpha)}$ is plotted against the
scaled variable $s/N^{D_s(\alpha)}$ in double logarithmic scales for
three different network sizes $N$ in Fig. \ref{DC}(a) for
$\alpha=2.25$ (in black) and $\alpha=4$ (in red) taking
$\epsilon=1/\sqrt{N}$. In Fig. \ref{DC}(b), the scaled avalanche area
distribution $P_a(\alpha,N)N^{D_a(\alpha)\tau_a(\alpha)}$ is shown for
the same values of $\alpha$ and $\epsilon$. Using the respective
values of the exponents, a good collapse of data are found to occur
for both $s$ and $a$ irrespective of the values of $\alpha$. Hence,
the FSS forms for $s$ and $a$ assumed in Eq. (\ref{psN}) are correct
over the wide range of $\alpha$ for a given $\epsilon$.

\section{Auto-correlation in toppling wave}
\begin{figure}[t]
\centerline{\hfill
  \psfig{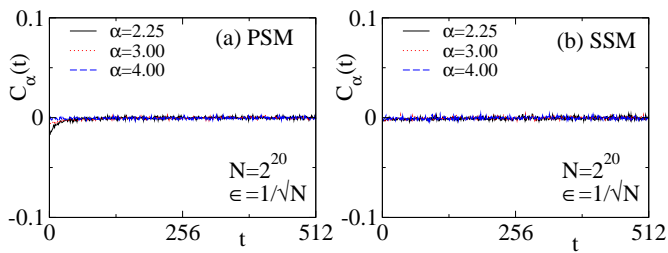}\hfill
}
\caption{\label{Ctw}(Colour online) Plot of $C_\alpha(t)$ against $t$
  for $\alpha=2.25$ (black solid line), $\alpha=3$ (red dotted line),
  and $\alpha=4$ (blue dashed line) in (a) for PSM and in (b) for
  SSM.}
\end{figure}

Since the model obeys FSS it is expected that there is no complete
toppling balance in PSM. Complete toppling balance refers to the fact
that the number of sands released by a toppled node is exactly equal
to the number of sands received by it when each of its neighbour nodes
topple once \cite{karmakarPRL05}. In PSM, if a node topples and gives
sand to its highest and lowest degree neighbours it is not necessarily
true that the toppled node is the highest or lowest degree node of any
of the nodes those received sands. Hence, it is expected that such a
toppling imbalance leads to the toppling waves generated from a fixed
critical node to be uncorrelated. A toppling wave is the number of
toppling during the propagation of an avalanche starting from a
critical node without further toppling of the same node
\cite{priezzhevPRL96,*ktitarevPRE00}. Each toppling of the critical
node creates a new toppling wave. The avalanche size $s$ can be
considered as $s=\sum_{j=1}^m s_j$, where $s_j$ is the size of the
$j$th wave and $m$ is the number of toppling waves in an
avalanche. The time auto correlation
\cite{demenechPRE00,*demenechPHYA02,*stellaPHYA01} in the toppling
waves on an SFN with given $\alpha$ is defined as
\begin{equation}
\label{cwave}
C_\alpha(t) = \frac{\langle s_{j+t}s_j\rangle - \langle
  s_j\rangle^2}{ \langle s_j^2\rangle - \langle s_j\rangle^2},
\end{equation}
where $t=1,2,\cdots$ and $\langle\cdots\rangle$ represents the time
average. $C_\alpha(t)$ is calculated for both PSM and SSM on a
network of size $N=2^{20}$ for several values of $\alpha$ taking
$\epsilon=1/\sqrt{N}$ and generating $10^6$ toppling waves for each
$\alpha$. $C_\alpha(t)$ is plotted against $t$ in Fig. \ref{Ctw} (a)
for PSM and in Fig. \ref{Ctw} (b) for SSM. It can be seen that for
both the models the values of $C_\alpha(t)$ are always zero for
different values of $\alpha$. Hence, the toppling waves are completely
uncorrelated as expected.

\section{Conclusion}
A two state sandpile model with preferential sand distribution is
constructed and studied on scale free network varying the degree
exponent $\alpha$. Due to the preferential constraint in the toppling
rule, the sand grains upon toppling of a critical node go to the
lowest and highest degree neighbour nodes. Such preferential sand
distribution leads to entirely different avalanche evolution than that
of the SSM in the scale free regime ($\alpha<4$) of the
network. Employing moment analysis, various exponents have been
estimated varying $\alpha$. For $\alpha\ge 4$, the exponents $\tau_s$
and $\tau_a$ become equal to $3/2$, the MF value whereas for
$\alpha<4$, $\tau_s<\tau_a<3/2$ and has a continuous dependence on
$\alpha$ in contrary to the results of the SSM in which
$\tau_s=\tau_a=3/2$ for the whole range of $\alpha$. The exponent
$\gamma_{sa}$ satisfies the scaling relation with $\tau_s$ and
$\tau_a$ within error bars. All the distribution functions of the
model satisfy FSS as there is no toppling balance and the time auto
correlation in the toppling wave is vanishingly small. The PSM,
sandpile model with preferential sand distribution, on SFN thus
belongs to a new universality class than that of the SSM in the scale
free regime of SFN.

\begin{comment}
All four exponents $\tau_s$, $\tau_a$ and $D_s$, $D_a$ are found to
increase as $\alpha$ varies from $2.05$ to $4$ and remain unchanged
there after {\em i.e.}  for $\alpha\ge 4$.  Though the exponents
$\tau_s$ and $\tau_a$ are found to be independent of $\epsilon$, the
capacity dimensions $D_s$ and $D_a$ depends on $\epsilon$.
\end{comment}

\bigskip
\noindent{\bf Acknowledgments:} This work is partially supported by
DST, Government of India through project
No. SR/S2/CMP-61/2008. Availability of computational facility,
``Newton HPC'' under DST-FIST project (No. SR/FST/PSII-020/2009)
Government of India, of Department of Physics, IIT Guwahati is
gratefully acknowledged.

\bibliographystyle{apsrev4-1}
\bibliography{santra_paper}

\end{document}